\documentclass[]{aa}
\usepackage{indentfirst}

\usepackage{txfonts}
\usepackage{graphicx}
\usepackage{xcolor}
\usepackage[]{hyperref}
\hypersetup{
    colorlinks,
    breaklinks=true,
    linkcolor={red},
    citecolor={blue},
    urlcolor={blue}
    }

\bibpunct{(}{)}{,}{a}{}{,}

\begin{document}
\title{Simulations of stellar winds from X-ray bursts.}

\subtitle{Characterization of solutions and observable variables.}

\author{ Y. Herrera \inst{\ref{UPC},\ref{IEEC}}
        \and
        G. Sala \inst{\ref{UPC},\ref{IEEC}}
        \and
        J. Jos\'e  \inst{\ref{UPC},\ref{IEEC}}.
        }

\institute{
    Departament de F\'isica, EEBE, Universitat Polit\`ecnica de Catalunya, c/Eduard Maristany 16, 08019 Barcelona, Spain.\label{UPC}
    \and
    Institut d’Estudis Espacials de Catalunya, c/Gran Capit\`a 2-4, Ed. Nexus-201, 08034 Barcelona, Spain. \label{IEEC}
}

\abstract { 
Photospheric radius expansion during X-ray bursts can be used to measure neutron star radii and help constrain the equation of state of neutron star matter. Understanding the stellar wind dynamics is important for interpreting observations, and therefore
stellar wind models, though studied in past decades, have regained interest and need to be revisited with updated data and methods.
}{
Here, we study the radiative wind model in the context of XRBs with modern techniques and physics input. We focus on characterization of the solutions and the study of observable magnitudes as a function of free model parameters.
}{
We implemented a spherically symmetric nonrelativistic wind model in a stationary regime, with updated opacity tables and modern numerical techniques. Total mass and energy outflows $(\dot M,\dot E)$ were treated as free parameters.
}{
A high-resolution parameter-space exploration was performed to allow better characterization of observable magnitudes.
High correlation was found between different photospheric magnitudes and free parameters. For instance, the photospheric ratio of gravitational energy outflow to radiative luminosity is directly proportional to the photospheric wind velocity.
}{
The correlations found here could help determine the physical conditions of the inner layers, where nuclear reactions take place, by means of observable photospheric values. Further studies are needed to determine the range of physical conditions in which the correlations are valid.
}

\maketitle

\section{Introduction} \label{sect:intro}

Type I normal (short) X-ray bursts (XRBs) are highly energetic and recurrent thermonuclear events occurring on the envelope of accreting neutron stars in binary systems where the secondary star is usually a main sequence star or red giant.
Most observed XRBs have short orbital periods in the range 0.2--15 hr. 
\footnote{Exceptions include GX 13+1 (592.8 hr), Cir X-1 (398.4 hr), and Cyg X-2 (236.2 hr), see \url{http://www.sron.nl/~jeanz/bursterlist.html} for an updated list of known galactic type I XRB systems.} 
As a result, the secondary star overfills its Roche lobe and mass-transfer ensues through the inner Lagrangian point (L1) of the system. 
The material stripped from the secondary has angular momentum such that it forms an accretion disk around the neutron star.
Viscous forces then progressively remove angular momentum from the disk forcing the material to spiral in and pile up on top of the neutron star.
The accreted material accumulates under mildly degenerate conditions, driving a temperature increase and the onset of nuclear reactions. 
As a result, a thermonuclear runaway occurs, generating a massive luminosity increase as well as nucleosynthesis of heavier elements, mostly around $A = 64$ (see, e.g., \cite{Jose2010,Fisker_2008,Woosley_2004}).

Type I XRBs are observationally characterized by a quick and sharp luminosity rise in about $1$ -- $10 \ \texttt{s}$, a burst duration ranging from $\sim 10$ to $100 \ \texttt{s}$, a total energy output of about $10^{39} \ \texttt{erg}$, and recurrence periods from hours to days. 
The luminosity can sometimes rise up to one thousand times the {persistent (accretion)} luminosity, although the typical increase is in the hundreds of times this latter, reaching values of the order of $10^{38} \ \texttt{erg/s}$. 
Another observable based on the light curve is the ratio between time-integrated persistent and burst fluxes, $\alpha$, typically in the range $\sim 40$ -- $100$,  corresponding to the ratio between the gravitational potential energy released by matter falling onto the neutron star during the accretion stage ($G M_\text{ns}/R_\text{ns} \sim 200 $ MeV per nucleon) and the nuclear energy generated in the burst (about 5 MeV per nucleon, for a solar mixture burned all the way up to the Fe-group nuclei). 
The presence of heavy elements can, theoretically, be detected (see \cite{WeinBildScha2006,ChangBildWasser2005,ChangMorBildWaser2006,BildChangPaer2003}) in the form of absorption features in the spectrum, which mostly lies in the X-ray range. 
For further information on XRBs, see \cite{StrohmBild2003, KeekZand2008, GalloMunHartChak2008,JoseBook2016}.

The mechanism powering XRBs bears a clear resemblance to that for classical novae, but unlike these latter, the high surface gravity of the neutron star prevents, in principle, the explosive ejection of material. 
However, the luminosity can approach or even exceed the Eddington limit, which may lead to the ejection of some material through a radiation-driven wind.

Stellar winds have been studied in different scenarios throughout most of twentieth century, and in a variety of forms (see \cite{Parker1965,Zytkow1972,CAK1975}). 
The simplest models assume spherical symmetry and stationary wind and can be broadly classified according to the main driving mechanism, namely gas pressure  or radiation, although magnetic fields can also play an important role. 
For a wind to be radiatively driven, high luminosity and high opacity must be present.

In the framework of neutron stars, several studies of radiation-driven winds have been performed since the early 1980s with varying hypotheses, approximations, and calculation techniques.
\cite{EbiHanaSugi1983}, \cite{Kato1983}, \cite{QuinnPacz1985} and \cite{JossMelia1984} all used nonrelativistic models with an approximated formula for opacity as a function of temperature. 
The models adopted different boundary conditions both at the photosphere and at the base of the wind envelope, as well as different treatments of the sonic point singularity (see Sect. \ref{sect:model}). 
Studies based on general relativistic models were performed by \cite{TuNoCa1986,PaczProsz1986}. 
In a more recent work, \cite{YuWeinberg2018} used MESA code \cite[see][]{MESA1} to perform a time-dependent hydrodynamic simulation of the wind envelope following a hydrostatic burst rise.

Advances in computational power and numerical techniques have allowed several studies since the 2000s to perform hydrodynamic simulations of XRBs with extended nuclear reaction networks (see \cite{Fisker_2008,Woosley_2004,Jose2010}). 
These studies have shown that XRBs synthesize a large variety of proton-rich nuclei. 
{The potential impact of XRBs on galactic abundances is still a matter of debate and relies on the high overproduction of some particular isotopes. 
It has been suggested that if a tiny fraction of the accreted envelope is ejected through radiation-driven winds, XRBs may potentially be the source of some light p-nuclei, such as $^{92,94}\text{Mo}$ and $^{96,98}\text{Ru}$ (see \cite{Schatz_1998, Schatz_2001}), which are systematically underproduced in all canonical scenarios proposed to date for the synthesis of such species.} 
However, it is still not clear whether XRBs contribute to the galactic abundances, because the strong gravitational pull of the neutron star prevents the direct ejection of matter, only potentially viable through a radiation-driven wind.
{Additionally, the study of XRBs wind can lead to a more accurate determination of neutron stars radii, and help constrain the  equation of state of neutron star matter (See \cite{Ozel2010,Steiner2010}).}
Thus, the interest in stellar winds in the context of XRBs has been renewed. 
For this reason, we proceeded to reanalyze these wind mechanisms, in an attempt to improve (with respect to some of the previously mentioned earlier studies from the 1980s) on some aspects of the input physics (e.g., updated opacity) and the numerical techniques (critical point treatment), and to extend the analysis of solutions, explore the parameter space in more detail, and characterize the observable magnitudes.

The paper is organized as follows. The basic wind model, input physics, and simulation setup are described in Sect. \ref{sect:model} and in Appendix \ref{sect:crit-substitution}. 
A full account of the results obtained from our simulations of radiative-driven winds is presented in Sect. \ref{sect:results}. Finally, the significance of our results and our main conclusions are summarized in Sect. \ref{sect:discuss}.

\section{Model, input physics, and initial setup} \label{sect:model}

Without any prior assumption on the temperature profile throughout the envelope, the spherically symmetric stationary wind equations constitute a set of nonlinear coupled differential equations. 
The boundary conditions are often given at more than one point and with implicit expressions. 
Another common feature is the appearance of a singular point where the wind becomes supersonic (see Sect. \ref{sect:boundaries}). 
Different approaches have been implemented to deal with the numerical difficulties involved in solving the equations close to this singular point (see \cite{Zytkow1972,JossMelia1984,PaczProsz1986}).

The simulations reported in this work rely on a stationary, nonrelativistic wind model with spherical symmetry.
Neutron stars have a typical radius of a few times their Schwarzschild radius, and so while general relativity corrections on gravity could be of importance close to its surface, they diminish rapidly in a wind envelope that extends hundreds of kilometers. 
{\cite{PaczProsz1986} showed that the main effects of general relativity on the wind structure can be relevant for solutions with lower mass outflows, resulting in a more extended (up to a factor x4) and cooler envelope. However, they rapidly become less significant for higher mass outflows ($> 5\times 10^{17}\texttt{g/s}$), giving corrections of only a few percent. We therefore leave the effects of general relativity for a follow-up work and study the Newtonian approach here as a stepping stone.}
The radiation-driven wind, treated as a fully ionized perfect gas, is assumed to be optically thick and in local thermal equilibrium (LTE) with radiation. 
{For a more detailed treatment of the optically thin regions and their impact on observables, see for example \cite{JossMelia1984}}

In this framework, the basic hydrodynamic equations for mass (\ref{Eq basic mass}), energy (\ref{Eq basic energy}), momentum conservation (\ref{Eq basic momentum}), and radiative energy transport (\ref{Eq basic transport}) become:
\begin{align}
  \label{Eq basic mass}  
  % &\text{Mass conservation:} & 
  && 
  \dot M &= 4 \pi r^{2} \rho v 
  \\
  \label{Eq basic energy}  
  % &\text{Energy conservation:} & 
  &&
  \dot E &= \dot M \left( \frac{v^{2}}{2} - \frac{G M }{r} + h \right) +L_{R} 
  \\
  \label{Eq basic momentum} 
  % &\text{Momentum conservation:} & 
  &&
  0 &= v \frac{\text{d}v}{\text{d}r} + \frac{G M }{r^{2}} + \frac{1}{\rho} \frac{\text{d}P}{\text{d}r} 
  \\
  \label{Eq basic transport}  
  % &\text{Energy transport:} & 
  &&
  \dfrac{\text{d}T}{\text{d}r} &= -\dfrac{3 \, \kappa \rho L_{R}}{16 \pi c a r^{2} T^{3}},
\end{align}
where the variables $r,T,v,\rho, L_\texttt{R} $ stand for the radial coordinate, temperature, wind velocity, gas density, and radiative luminosity, respectively. 
The total pressure $P$, and specific enthalpy $h$, include the contributions of gas and radiation, in the form: 
\begin{align}
  &&
  P{(\rho,T)} &= \frac{\rho k T}{\mu m_A} +\frac{a T^{4}}{3} 
  & \text{and}
  &&
  h{(\rho,T)} &= \frac{5}{2} \frac{k T}{\mu m_A} + \frac{4}{3} \frac{a T^{4}}{\rho}.
  \label{eq: Pressure and enthalpy}
\end{align}

The opacity $\kappa{(T,\rho)}$ must be provided externally either from a radiation transport theoretical model, phenomenological relations, or experimental values. 
Opacity can have different sources depending on the microscopic processes involved; in general, these depend on photon frequency. 
The tables calculated by OPAL (\cite{OPAL1996}) take these aspects into account to give a Rosseland mean opacity as a function of temperature, density, and composition of the gas only. 
These tables have become the standard source for stellar opacity in recent decades and are used in the present work. 
\footnote{{Opacity tables employed are limited to $\log T < 8.7$. For higher temperatures, opacity was extrapolated using a formula introduced by \cite{Paczynski1983}.}}

The total mass and energy outflows $(\dot M, \dot E)$ arise as integration constants from the mass and energy conservation laws and can be determined by imposing conditions at the base of the wind. 
Therefore, $\dot M, \dot E $ are model parameters. 
During an X-Ray burst these will vary in time, but are assumed to be  approximately constant across the wind for a fixed time.
The mass of the neutron star core $M$ is considered fixed and constitutes the only relevant source of gravity (the contribution of the envelope mass can be neglected). 
Lastly the mean molecular mass $\mu$ is a function of the mass fractions $X_i$ of the different species present in the envelope.
The rest of the symbols have their usual meaning: $G$ for gravitational constant, $c$ for speed of light, $k$ for Boltzmann constant, $a = \frac{4\sigma}{c}$ for radiation energy density constant ($\sigma$ is the Stefan-Boltzmann constant) and $m_A$ for atomic mass unit.
 
By differentiating the mass conservation equation \ref{Eq basic mass}, for a constant $\dot M$, one can obtain:
\begin{align}
  &&
  0 = 2 \frac{ \text{d}r}{r} + \frac{\text{d}\rho}{\rho} + \frac{\text{d}v}{v}.
  \label{eq: Mass conservation differential}
\end{align}
One can also calculate, from eq. \ref{eq: Pressure and enthalpy}, the pressure differential appearing as the last term in the momentum equation \ref{Eq basic momentum}:
\begin{align}
  &&
  \frac{\text{d}P}{\rho} 
  &=  \frac{kT}{\mu m_A}  \left( \frac{\text{d}T}{T} + \frac{\text{d}\rho}{\rho} \right) 
  + \frac{4}{3} \frac{a T^4}{\rho} \frac{\text{d}T}{T},
\end{align}
and by using eq. \ref{eq: Mass conservation differential} to replace the density differential: 
\begin{align}
  &&
  \frac{\text{d}P}{\rho} &= - \frac{kT}{\mu m_A} \left( \frac{2 \text{d}r}{r}  + \frac{\text{d}v}{v} \right) + \left( \frac{kT}{\mu m_A} + \frac{4}{3} \frac{a T^4}{\rho} \right) \frac{\text{d}T}{T}.
 \end{align}
 
 After some regrouping, both differential equations \ref{Eq basic momentum} and \ref{Eq basic transport} can be put in the form:
\begin{align}
  \label{eq: momentum diff}
  \left( v^2 - \frac{k T}{\mu m_A} \right) \frac{\text{d}v}{v} + \left( \frac{k T}{\mu m_A}  +\frac{4}{3} \frac{a T^4}{\rho} \right) \frac{\text{d}T}{T} + \left( \frac{G M }{r} - 2 \frac{k T}{\mu m_A} \right) \frac{\text{d}r}{r} &= 0 \\
  \label{eq: transfer diff}
  \left( \frac{4}{3} \frac{a T^4}{\rho} \right) \frac{\text{d}T}{T} + \left( \frac{G M }{r} \Gamma \right) \frac{\text{d}r}{r}&= 0 ,
\end{align}
where we introduced the luminosity ratio $\Gamma = L_\texttt{R}/L_\texttt{Edd} = \frac{\kappa L_R}{4\pi c GM}$ ($L_\texttt{Edd}$ is the local Eddington luminosity)
and the density and radiative luminosity can be obtained using the mass and energy conservation equations \ref{Eq basic mass} and \ref{Eq basic energy}:
\begin{align}
    &&
    \label{eq:density from mass conservation}
    \rho &= \frac{\dot M }{4 \pi r^{2}  v} 
    \\
    &&
    \label{eq:luminosity from energy conservation}
    L_R &= \dot E - \dot M  \left( \frac{v^2}{2} - \frac{G M }{r}  + \frac{5}{2} \frac{k T}{\mu m_A}  +\frac{4}{3} \frac{a T^4}{\rho} \right).
\end{align}
These constitute a system of four equations, two algebraic (\ref{eq:density from mass conservation} and \ref{eq:luminosity from energy conservation}) and two first-order ordinary differential equations (\ref{eq: momentum diff} and \ref{eq: transfer diff}) that can be solved for $v, T, \rho, L_R$ as functions of the independent variable $r$.

\subsection{Boundary conditions} \label{sect:boundaries}

Two boundary conditions are required to solve the system of two first-order differential equations. The first condition is given by the definition of the photosphere, where the gas temperature is equal to the effective temperature $T_\texttt{eff}$, and can be written as:
\begin{align}
  &&
  T_\text{ph} = T_\text{eff} &= \left( \frac{L_\text{R,ph}}{\pi a c r^2_\text{ph}}  \right)^{\tfrac{1}{4}},
  \label{temperature condition}
\end{align}
which shall be referred to as ``temperature condition''. Hereafter, sub-index ``ph'' denotes evaluation at the photosphere.

The location of the photosphere in radius is not known a priori so this adds an extra unknown parameter. This is usually given by the point at which  the optical depth is about unity, which would require the integration of the atmosphere above the photospheric surface. In the case of an extended atmosphere it is usual to take a locally defined effective optical depth $\tau^* = \kappa \rho r $ that takes a minimum at the photosphere with an approximate value of $\tau^*_\text{ph} \simeq 3 $ \cite[see][]{Kovetz1998}. \cite{KatoHachisu1994} showed that this minimum usually corresponds to the point where this effective optical depth takes a value closer to $\tau^*_\text{ph} \gtrsim 8/3$. Accordingly, we adopt the following ``optical condition'':
\begin{align}
  &&
  \tau^*_\text{ph} & =  \frac{8}{3} \simeq 2.67 .
  \label{optical condition}
\end{align}
These conditions, \ref{temperature condition} and \ref{optical condition}, when replaced in the radiation diffusion equation \ref{Eq basic transport} give rise to a condition for the temperature gradient at the photosphere:
\begin{align}
  &&
  \frac{\text{d}\log T}{\text{d}\log r}\bigg|_\text{ph} &\simeq - \frac{1}{2} ,
  \label{gradient condition}
\end{align}
which can be used alternatively instead of either the optical or temperature conditions.

The second boundary condition arises from analyzing the topology of solutions to the momentum equation. From equation \ref{eq: momentum diff}, the wind velocity gradient can be expressed as: 
\begin{align}
  &&
  \frac{r}{v} \frac{\text{d}v}{\text{d}r} &= - \frac{ \left( \frac{k T}{\mu m_A}  +\frac{4}{3} \frac{a T^4}{\rho} \right) \frac{r}{T} \frac{\text{d}T}{\text{d}r} + \left( \frac{G M }{r} - 2 \frac{k T}{\mu m_A} \right) }{\left( v^2 - \frac{k T}{\mu m_A} \right)} ,
\end{align}
from where it is clear that the momentum equation presents a singularity when:
\begin{align}
  &&
  v^2 &= \frac{k T}{\mu m_A} ,
  \label{singularity condition}
\end{align}
that is, when the wind velocity is equal to the local isothermal sound speed (hereafter, singularity condition).
If the conditions were such that this singular value were met, in order to have a nonsingular physical value for the velocity gradient we must simultaneously require the following ``regularity condition'':
\begin{align}
  &&
  \left( \frac{k T}{\mu m_A}  +\frac{4}{3} \frac{a T^4}{\rho} \right) \frac{r}{T} \frac{\text{d}T}{\text{d}r} + \left( \frac{G M }{r} - 2 \frac{k T}{\mu m_A} \right)&= 0 ,
  \label{regularity condition}
\end{align}
so that the quotient is finite. 
A solution may or may not pass through this ``critical point'' a priori, but the only physically acceptable solution with a wind velocity that approaches zero towards the bottom of the envelope and always increases outwards is the one that indeed does pass through this point. Furthermore, the wind becomes supersonic above this critical point (also known as the sonic point). 
It is not known a priori where this critical point will lie, but together conditions \ref{singularity condition} and \ref{regularity condition} provide both the extra condition needed to locate it and the second boundary condition for the integration constant. We use the subscript ``cr'' to refer to variables evaluated at the critical sonic point.

We therefore have four unknown parameters $v_\text{cr}, r_\text{cr}, T_\text{ph},r_\text{ph}$ with four conditions to solve them: two at the photosphere (temperature and optical conditions) and two at the critical point (singularity and regularity conditions). The system  constitutes a two-point boundary-value problem with nonlinear first-order differential equations. Standard numerical methods for dealing with them are the relaxation method and the shooting method.

{However, the resulting solutions may not always reach physical values compatible with those of a neutron star at the wind base. For instance, a high temperature may be found at radii much larger than those of typical neutron stars; on the other hand, velocity may not be realistically low enough (or temperature/density high enough) when reaching a typical neutron star radius. 
Previous works relied on simplified assumptions for the wind base conditions, such as for example a fixed value of density or temperature, or an energy generation rate given by integration of simple nuclear models.
We find that these choices for a wind base may not always be compatible with some of the recent and more realistic hydrodynamic simulations for X-ray bursts (see \cite{Jose2010}).
}

{In order to accurately determine the conditions at the wind base, hydrodynamic simulations of the nuclear burning layers are likely required, but this is clearly out of the scope of this paper. Here, we   search for a condition for the wind base that is realistic for a neutron star undergoing an X-ray burst, while at the same time leave some margin for the natural variability of the physical conditions. Setting a range of values for all the physical variables seems like a good approach at first, but the choice of several restriction values seems arbitrary and turned out to not always be compatible with self-consistent solutions. For instance, a particular choice of maximum velocity and minimum temperature at the wind base sometimes ended up giving solutions that were not deemed stationary (according to the characteristic time criterion explained in Sect. \ref{sect:method}). Instead, we chose a criterion that is physically based, gives similar wind structure for all solutions, and reduces the amount of variables to be restricted, thus reducing arbitrariness, while at the same time still being compatible with the expected conditions for X-ray bursts. We set the wind base at the innermost point where the radiation pressure gradient becomes larger than the gas pressure gradient, that is where:
} 
\begin{align}
  & & \label{Wind Base condition}
  \nabla P_\texttt{R} &\geq \nabla P_\texttt{g} 
  & \texttt{or} & &
  \frac{dP_\texttt{g}}{dP_\texttt{R}} &\leq 1 .
\end{align}
{Above this point, radiation pressure becomes more important than gas pressure as the driving mechanism.
}
{
Furthermore, in all solutions explored, the velocity gradient term $v \frac{dv}{dr}$ in the momentum conservation equation \ref{Eq basic momentum} is still negligible where the equality in equation \ref{Wind Base condition} is met, and so the physical conditions can be considered to approximately match those of a static envelope. It can also be shown that at this point $\Gamma \simeq \frac{1}{2}$. 
}

{
Solutions that meet the wind base condition in Eq. \ref{Wind Base condition} at radii compatible with possible neutron stars ($7 - 20 \texttt{ km}$) are then selected and considered as possible candidates for further study. In Sects. \ref{sect: wind profiles} and \ref{sect:WiMPS} we discuss the validity of these solutions for describing XRBs.
}

\subsection{Numerical procedure} \label{sect:method}

The integration method used to solve the differential equations is the adaptive-step Runge-Kutta (RK45). 
In order to deal with the two-point boundary-value problem (photosphere and critical point) we use the shooting method as follows. 
Integration starts at a critical point with radius $r_\text{cr}$ and velocity $v_\text{cr}$ satisfying both singularity and regularity conditions for a temporarily assumed $T_\text{cr}$. 
Integration proceeds outwards until the optical photospheric condition is met $\left(\kappa \rho r \simeq 8/3\right)$.
There we parametrize a distance to the photospheric temperature condition with a value $\phi$:
\begin{align}
  &&
  \phi &= \frac{T - T_\text{eff}}{T + T_\text{eff}},
\end{align}
and then use a suitable root-finding method for $\phi$ in terms of the chosen starting value for $T_\text{cr}$. 
The numerical difficulties of starting the integration from the critical point were solved through a change of variables that cancels the singular denominator in the velocity gradient (see Appendix \ref{sect:crit-substitution}).

The integration from the critical point to the photosphere corresponds to the region where the wind is supersonic. Once a supersonic profile is found, we go back to the critical point and proceed integrating inwards.
The integration is stopped either when the temperature exceeds a given limit ($ T > 10^{10} \ \texttt{K}$) or the Schwarzschild radius is reached.  

The solutions are checked for some hypothesis consistency and classified accordingly, in a similar fashion to that in \cite{QuinnPacz1985}, as follows.
In order for the wind to be considered thick, most of the acceleration must happen at high optical depth. It is enough to require that the sonic point lies at $\tau >> 1 $. Solutions for which the effective optical depth of the critical point is too small ($\tau_c^*  < 10 $) are deemed as optically ``thin''. 
The stationary hypothesis is also checked by comparing the dynamical characteristic times $\Delta t $ above and below the critical point, where
\begin{align}
  &&
  \Delta t &= \int_{r_1}^{r_2} \frac{dr}{v},
\end{align}
which is an estimate of the travel time of a fluid element between arbitrary radii $r_1$ and $r_2$. 
For the solutions to be considered stationary, $\Delta t_\texttt{above} << \Delta t_\texttt{below}$, which by use of mass conservation translates into the $\Delta m_\texttt{above} << \Delta m_\texttt{below}$ requirement for the envelope mass. Solutions with $ \Delta m_\texttt{above} > 0.1 \ \Delta m_\texttt{below}$ were considered to be nonstationary. We note that the value of $\Delta m_\texttt{below}$ depends on the choice of the inner boundary for the wind base, and the classification of a particular solution may change with a different cut-off for the integration inwards or when imposing the nuclear burning shell conditions. In every solution found, the photospheric velocity never exceeded a few percent of the speed of light ($ v_\text{ph} \lesssim 0.03\ c $), and so the nonrelativistic hypothesis is safe. The hypothesis consistency criteria mentioned above are fairly common and can be found in several works (see e.g. \cite{JossMelia1984}, \cite{KatoHachisu1994}).

The procedure was repeated for different values of the wind parameters $( \dot M, \dot E)$. In all the simulations reported in this work, a neutron star mass of $M = 1.4 \ \texttt{M}_\odot$ has been adopted, together with an envelope composition given by $X=0$, $Y= 0.9$, $Z = 0.1$. {Although recent simulations of XRBs (\cite{Jose2010,Fisker_2008,Woosley_2004}) show higher metallicities, the aforementioned values were the highest available on the opacity tables used. More realistic values will be implemented in future work as opacity data become available.}

\section{Results} \label{sect:results}

Simulations were run for a wide range of values in the wind parameter space $(\dot M, \dot E)$, exploring over 1000 points. We normalized the energy output $\dot E$ in terms of a constant Eddington luminosity 
\begin{align}
  &&
  L_\text{o} &= \frac{4\pi cGM}{\kappa_\text{o} (1+X_\texttt{H})},
\end{align}
where $\kappa_\text{o} = 0.2 \  \texttt{cm}^{2} \texttt{g}^{-1}$ is the electron scattering opacity, and $X_\texttt{H}$ is the hydrogen abundance. 
For a neutron star with $M = 1.4 \ \texttt{M}_\odot$ and no hydrogen in the wind envelope, $L_\text{o} \simeq 3.51 \times 10^{38} \ \texttt{erg }\texttt{s}^{-1}$.

\subsection{Wind profiles} \label{sect: wind profiles}
\begin{figure}
    \centering 
    \includegraphics[keepaspectratio=true,width=8cm,clip=true,trim=0pt 70pt 0pt 28pt]{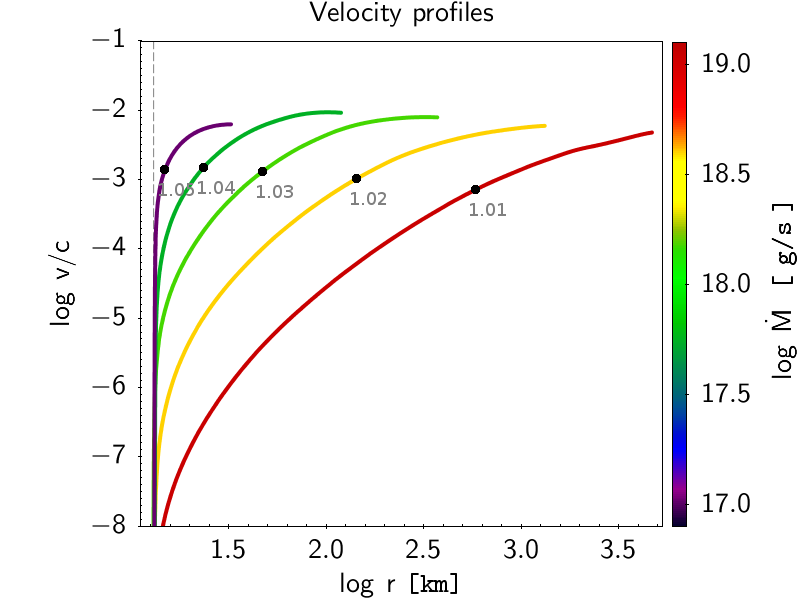}\\
    \includegraphics[keepaspectratio=true,width=8cm,clip=true,trim=0pt 70pt 0pt 28pt]{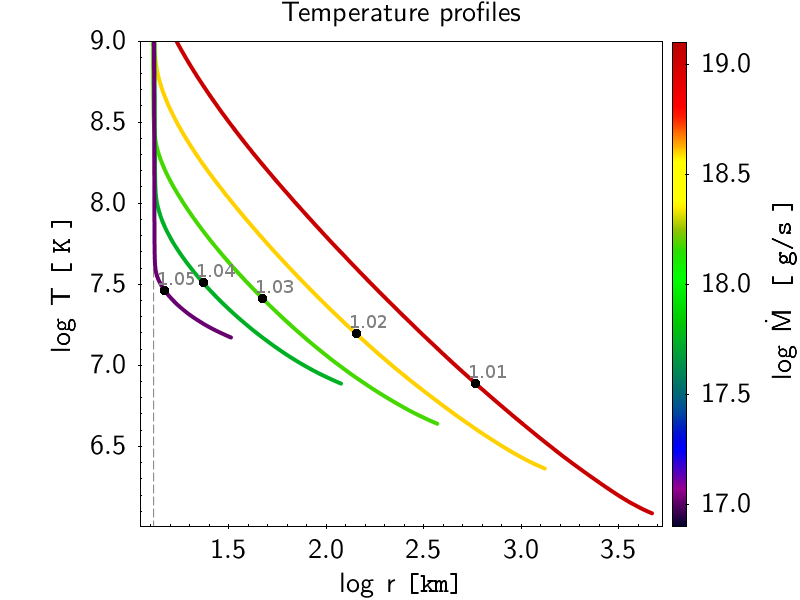}\\
    \includegraphics[keepaspectratio=true,width=8cm,clip=true,trim=0pt 70pt 0pt 28pt]{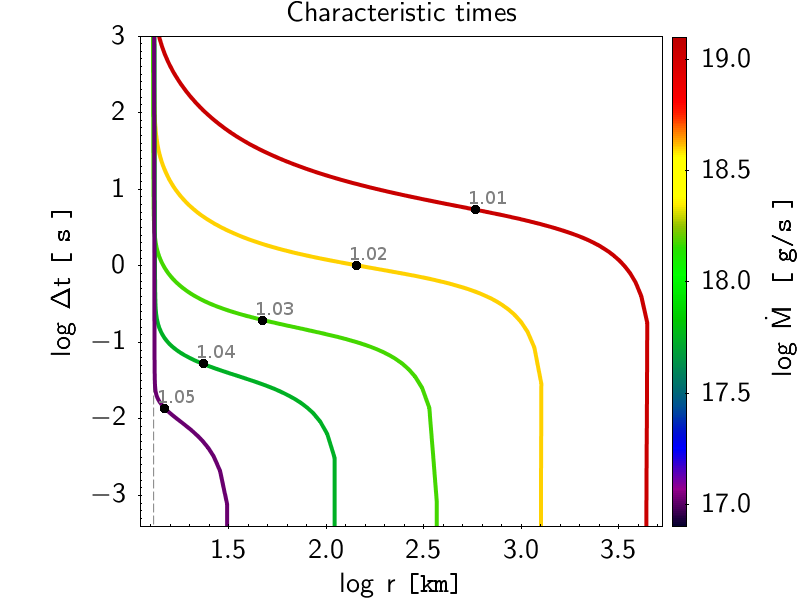}\\
    \includegraphics[keepaspectratio=true,width=8cm,clip=true,trim=0pt 0pt 0pt 28pt]{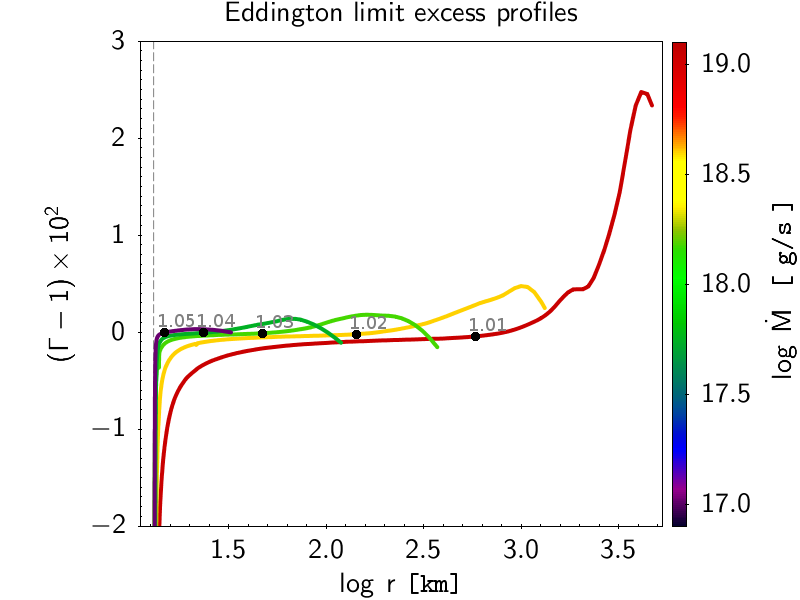}\\
    \caption{{Wind profiles obtained that are compatible with a neutron star radius of $13 \texttt{ km}$ (vertical dashed line), with different values of parameters ($\dot M, \dot E$).
    Values of mass outflow $\dot M$ are indicated by the line color, and values of energy outflow $\dot E$, in units of $L_\texttt{o}$ (see text), are indicated with labels next to a black circle ($\bullet$), which marks the location of the critical sonic point in each curve. Top to bottom: Velocity, temperature, characteristic time, and luminosity ratio $\Gamma$, all presented as a function of radius. All curves end at the photosphere. } }
    \label{fig:Profiles} 
 \end{figure}

A suite of different wind profiles is displayed in Fig. \ref{fig:Profiles}. 
{Each plot presents a selection of solutions that have a wind base compatible with a neutron star radius of $R_\texttt{ns} = 13 \texttt{ km}$. 
% (marked by a vertical dashed gray line). 
% Line color indicates mass outflow $\dot M$, critical point is marked in each profile with a black dot ($\bullet$), and energy outflow value $\dot E$ (in units of $L_o$) is tagged next to each profile. 
The first two plots in Fig. \ref{fig:Profiles} are solutions for velocity and temperature, respectively. The third plot is a profile of characteristic time $\Delta t$, which is the time it takes for a fluid element to reach the photosphere from a given radius. The bottom plot is the luminosity ratio $\Gamma = L_R / L_\texttt{Edd}$ (where $L_\texttt{Edd}$ is the local Eddington luminosity).}

{The general pattern observed is that solutions with smaller energy outflows are characterized by larger radii, but also a higher mass output and longer characteristic timescales. This is due to the fact that a
fluid element requires longer distances to accelerate to a sufficiently high velocity
% acceleration of a fluid element takes longer distances 
when its mass is larger or the available energy is smaller. 
Profiles with higher mass outflow (lower energy outflow) may not be suitable for describing short XRBs, since the total characteristic time (from wind base to photosphere) would be larger than the typical burst duration. }

Another significant behavior is that the radiative luminosity starts very low in comparison with the local Eddington limit at low radii, suggesting that the wind is driven by gas pressure rather than radiation in the inner regions. The luminosity ratio $\Gamma = {L_\text{R}}/{L_\text{Edd}}$ then rises sharply, before reaching the critical point and flattens very close to unity after that, remaining flat for the rest of the supersonic profile, except for some minor bumps ( about $ 3 \%$ maximum excess) in some models close to the photosphere.

\subsection{Parameter space exploration} \label{sect:WiMPS}

\begin{figure}
    \centering
    \includegraphics[keepaspectratio=true,width=8cm,clip=true,trim=0pt 80pt 0pt 40pt]{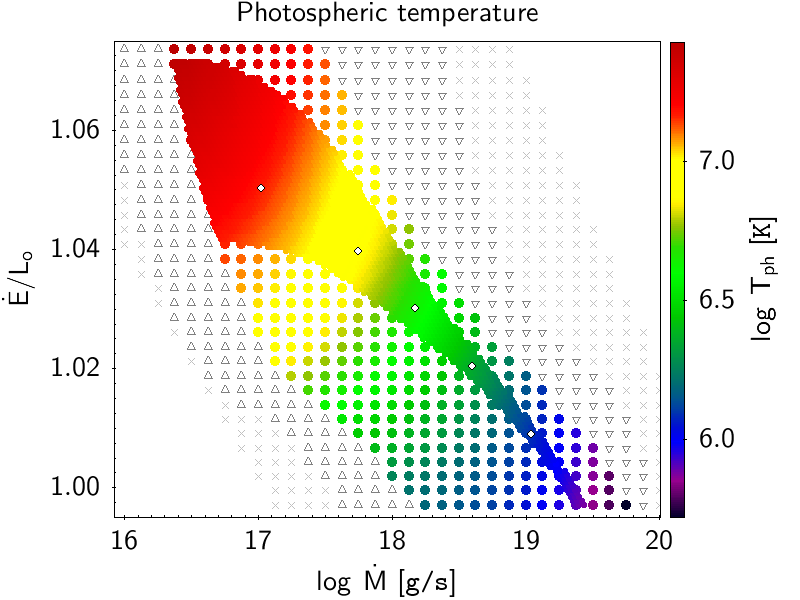} \\
    \includegraphics[keepaspectratio=true,width=8cm,clip=true,trim=0pt 80pt 0pt 40pt]{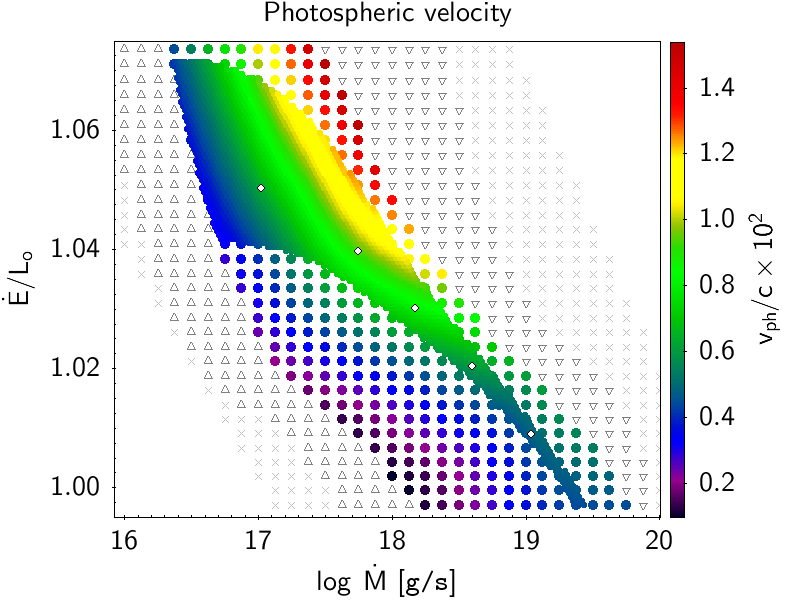}\\
    \includegraphics[keepaspectratio=true,width=8cm,clip=true,trim=0pt 80pt 0pt 40pt]{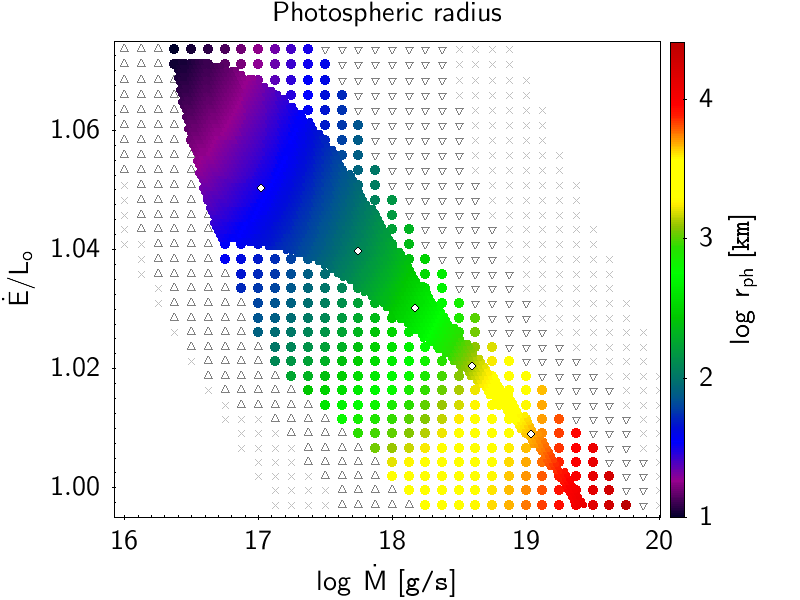}\\
    \includegraphics[keepaspectratio=true,width=8cm,clip=true,trim=0pt 0pt 0pt 40pt]{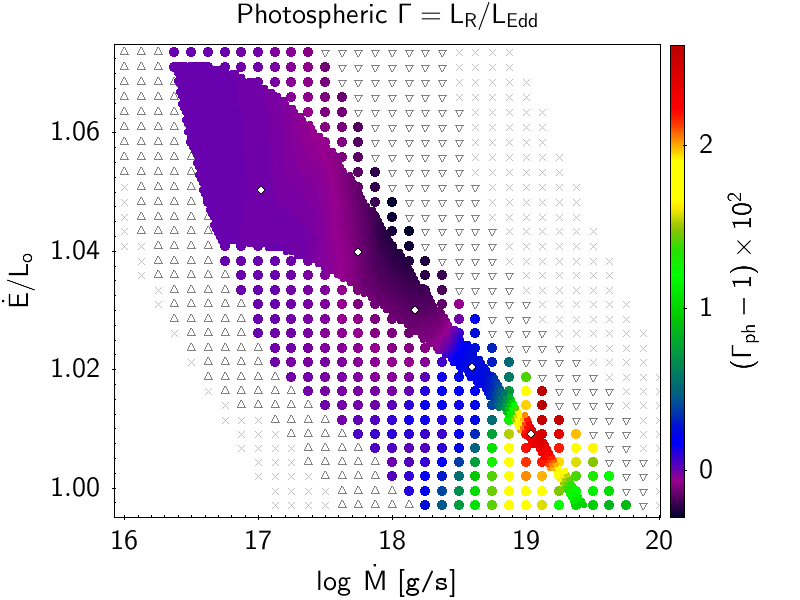}
    \caption{Wind parameter space sweep. Color coded values of photospheric magnitudes for different values of parameters ($\dot M, \dot E$). From top to bottom: Temperature, velocity, radius, and luminosity ratio $\Gamma$. Points marked with up or down triangles ($\triangle / \triangledown$) correspond to ``thin'' and nonstationary solutions, respectively. Points marked with filled circles ($\bullet$) are self-consistent solution candidates. The $\times$ symbol denotes no solutions found for the given boundary conditions. {The fully colored area marks solutions whose wind base is compatible with possible neutron star radii ($7 - 20 \texttt{ km}$), and white diamonds inside of it ($\diamond$) mark selected solutions plotted in Fig. \ref{fig:Profiles}.}}
    \label{fig:WiMPS-Photo} 
\end{figure}

\begin{figure}
    \centering
    \includegraphics[keepaspectratio=true,width=8cm,clip=true,trim=0pt 80pt 0pt 40pt]{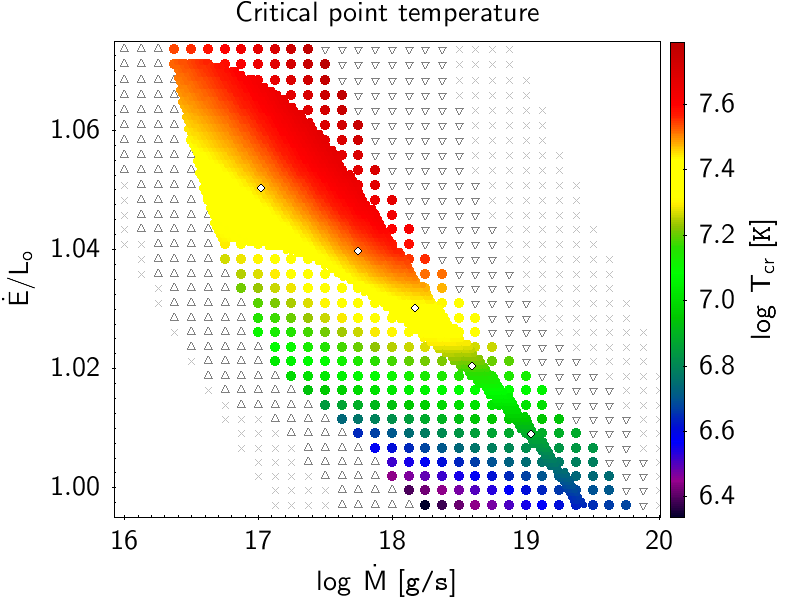}\\
    \includegraphics[keepaspectratio=true,width=8cm,clip=true,trim=0pt 80pt 0pt 40pt]{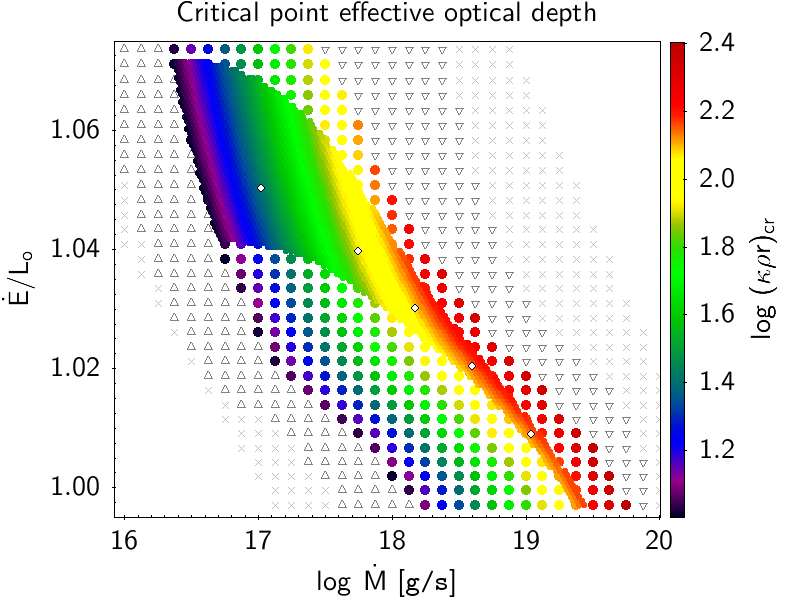}\\
    \includegraphics[keepaspectratio=true,width=8cm,clip=true,trim=0pt 80pt 0pt 40pt]{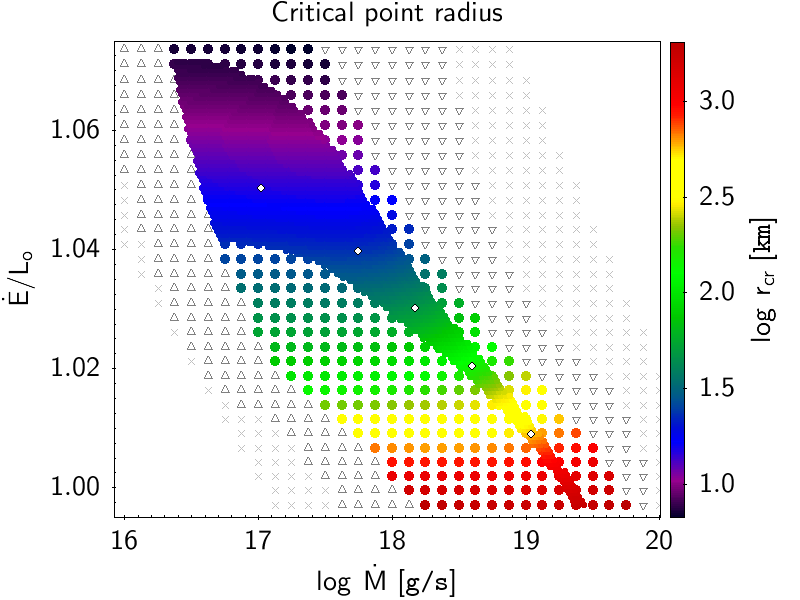}\\
    \includegraphics[keepaspectratio=true,width=8cm,clip=true,trim=0pt 0pt 0pt 40pt]{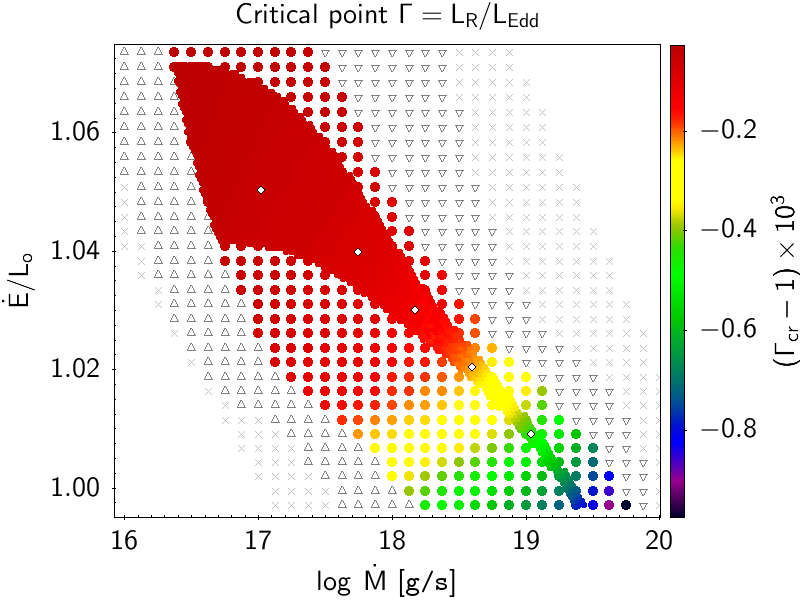}
    \caption{Same as Fig. \ref{fig:WiMPS-Photo}, but for values at the critical (sonic) point, with the exception of the velocity plot that has been replaced by an effective optical depth $\tau^* = \kappa \rho r$ plot.}
    \label{fig:WiMPS-Crit} 
\end{figure}

\begin{figure}
    \centering
    \includegraphics[keepaspectratio=true,width=8cm,clip=true,trim=0pt 80pt 0pt 40pt]{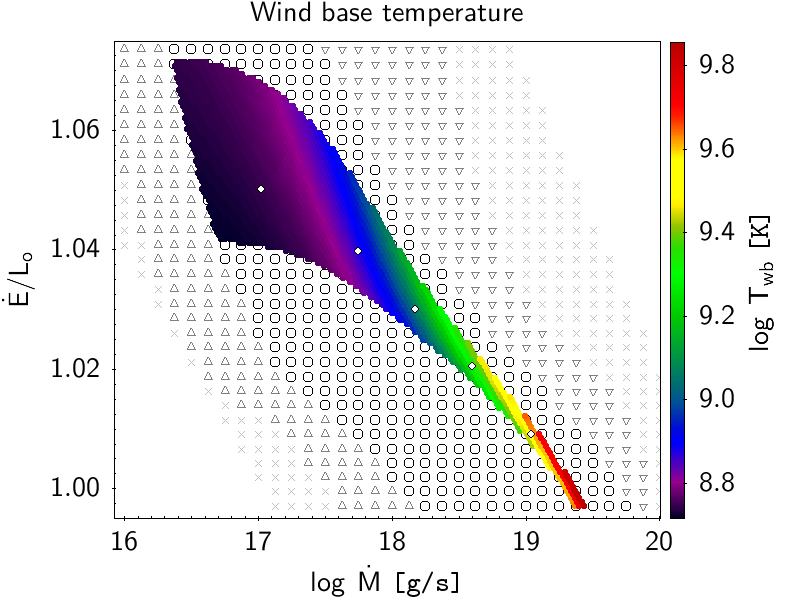}\\
    \includegraphics[keepaspectratio=true,width=8cm,clip=true,trim=0pt 80pt 0pt 40pt]{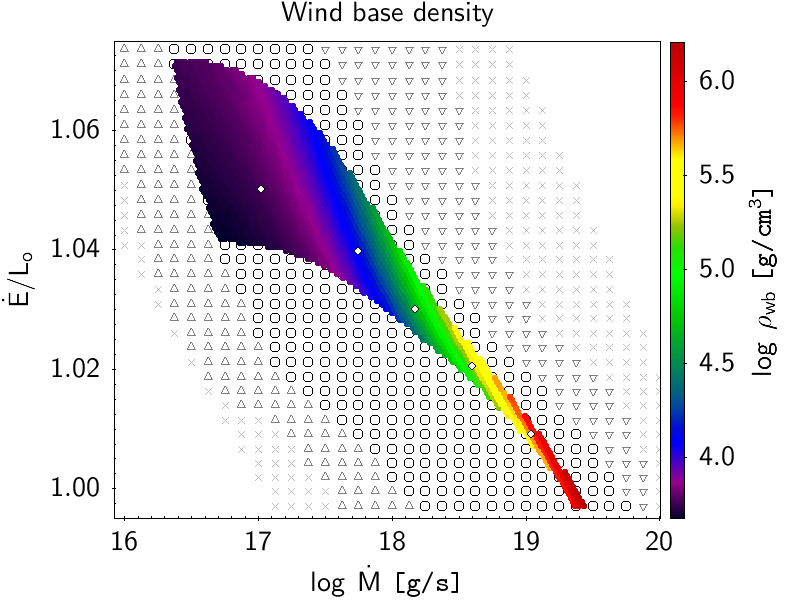}\\
    \includegraphics[keepaspectratio=true,width=8cm,clip=true,trim=0pt 80pt 0pt 40pt]{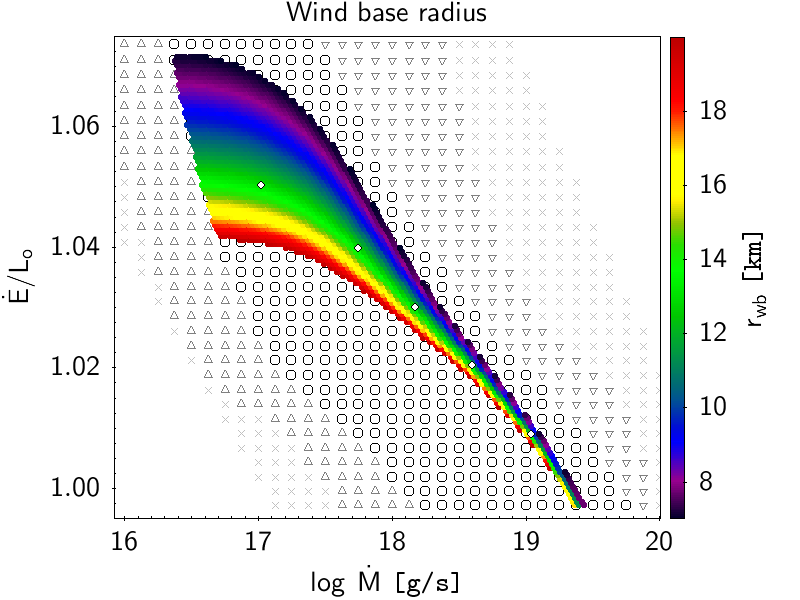}\\
    \includegraphics[keepaspectratio=true,width=8cm,clip=true,trim=0pt 0pt 0pt 40pt]{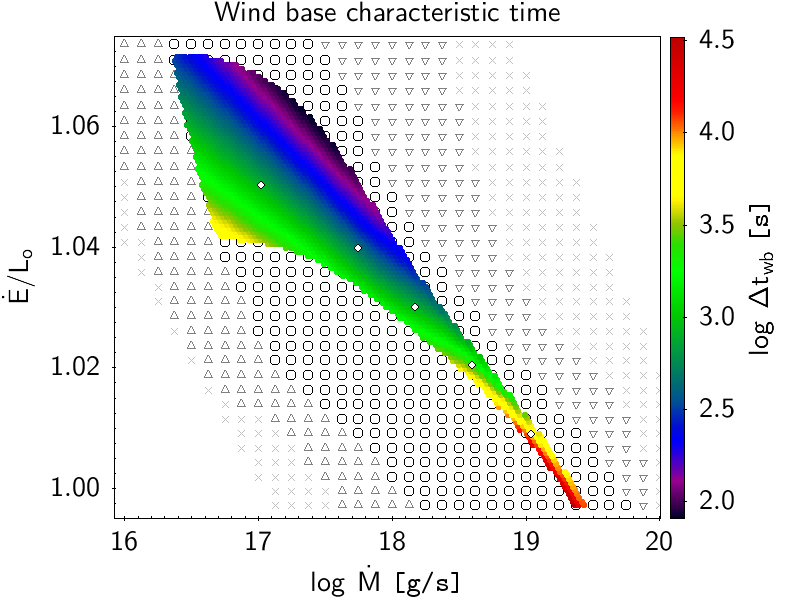}
    \caption{{Same as Fig. \ref{fig:WiMPS-Photo}, but for values of temperature, density, radius, and characteristic time $\Delta t $ at the wind base. Self-consistent solutions outside the area of compatibility with neutron stars are not colored here in order to better resolve values in the area of interest.}}
    \label{fig:WiMPS-Base} 
\end{figure}

{Figures \ref{fig:WiMPS-Photo} to \ref{fig:WiMPS-Base} show photospheric, critical point and wind base values, respectively, as a function of model parameters $ (\dot M, \dot E)$. Temperature and radius are plotted in all three figures. Luminosity ratio $\Gamma $ is plotted in photosphere and critical point figures ($\Gamma_\texttt{wb} \simeq 0.5$ is constant). Wind velocity is only shown for the photosphere, because at the critical point, $v^2_\texttt{cr} \sim T_\texttt{cr}$, and wind base values are negligible with values smaller than a few centimeters per second. Other relevant plots presented are effective optical depth $\tau^*$ for critical point only, and wind base density and characteristic time.}
% The points for which no solution was found for the boundary-value problem are marked with a faint $\times$.  As mentioned before, if a solution's critical point did not lie above a high effective optical depth $\tau^* = \kappa \rho r > 10 $ the solution is considered optically ``thin''; these are marked with an up-triangle ($\vartriangle$). Solutions not complying with the steadiness criteria are marked with a down-triangle ($\triangledown$). Self-consistent (acceptable) solutions are indicated with circles ($\bullet$). {The fully colored high resolution area indicates solutions whose wind base lies in a radial range compatible with a neutron star ($7 -20 \texttt{ km}$). White diamonds ($\diamond$) inside this area mark the selected solutions for $r_\texttt{ns} \simeq 13 \texttt{ km} $ plotted in Fig \ref{fig:Profiles}.} 
% In each plot, the color coding indicates the value of the physical magnitude of interest.
All figures show both the acceptable (self-consistent) solutions and  those discarded for being either optically thin or nonstationary solutions, according to the criteria explained in Sect. \ref{sect:method}. The subset of acceptable solutions that are compatible with possible neutron stars is also indicated, as well as the solutions plotted in Fig. \ref{fig:Profiles}. 
We note that the criterion for stationary wind depends on the ratio of envelope mass above and below the critical point, and thus the 
choice of the wind base affects this classification.

We observe a clear pattern: as energy output $\dot E$ increases and mass output $\dot M $ decreases, photospheric temperature rises, in the exact opposite fashion to the photospheric radius, which decreases. We  study this correlation further below in Sect. \ref{sect:correlations}. The photospheric wind velocity gradient with respect to parameters seems to be roughly orthogonal to that of temperature or radius, only achieving a few percent of speed of light, and peaking outside the self-consistent solutions zone. The photospheric radiative luminosity is very close to local Eddington luminosity, as shown by the $\Gamma$ plot, except for low-energy and high-mass outflows, where it peaks at up to a $3\%$ excess. This is due to a small bump in opacity tables that lowers the local Eddington luminosity.

As for the critical point (Fig. \ref{fig:WiMPS-Crit}), no obvious relationship between temperature and radius appears. The change in critical radius seems to go mostly inversely with $\dot E$, while the change in critical temperature seems to have a bigger component in the $\dot M$ direction, {especially} at high $\dot E$. By definition of critical point, a critical velocity plot would give the same information as the temperature one, instead we show a plot for effective optical depth. The critical point lies at higher optical depth with bigger mass outflow, hinting at a stretching of the envelope as mentioned above. The radiative luminosity is always close to the local Eddington limit, but just slightly lower, by no more than $\sim 0.1\%$.

{ The temperature and density values at the wind base (as defined by equation \ref{Wind Base condition}) are shown in Fig \ref{fig:WiMPS-Base}. Their values are in a reasonable range for XRBs conditions (See \cite{Jose2010,Fisker_2008,Woosley_2004}) in the area of radial compatibility with neutron stars. However, not all of the models in this area have wind base characteristic times (i.e. the time required for a fluid element to reach the photosphere from the wind base) that are well suited for describing short XRBs, whose duration is of the order of few hundreds of seconds. These  models could still represent winds in long-duration XRBs. 
{Nevertheless, it is possible that the high values of $\Delta t$ found are a result of the choice of inner boundary condition, as mentioned  in Sect. \ref{sect:method}. In the profiles shown in Fig. \ref{fig:Profiles}, the velocities quickly go to zero at the wind base, and therefore wind material spends nearly all its outflow time just above the wind base. This is probably an artifact of the time-independent treatment, which may no longer be accurate close to the inner boundary. 
}
}

\subsection{Photospheric correlations} \label{sect:correlations}

\begin{figure}
    \centering
    \includegraphics[keepaspectratio=true,width=8cm,clip=true,trim=0pt 0pt 0pt 0pt]{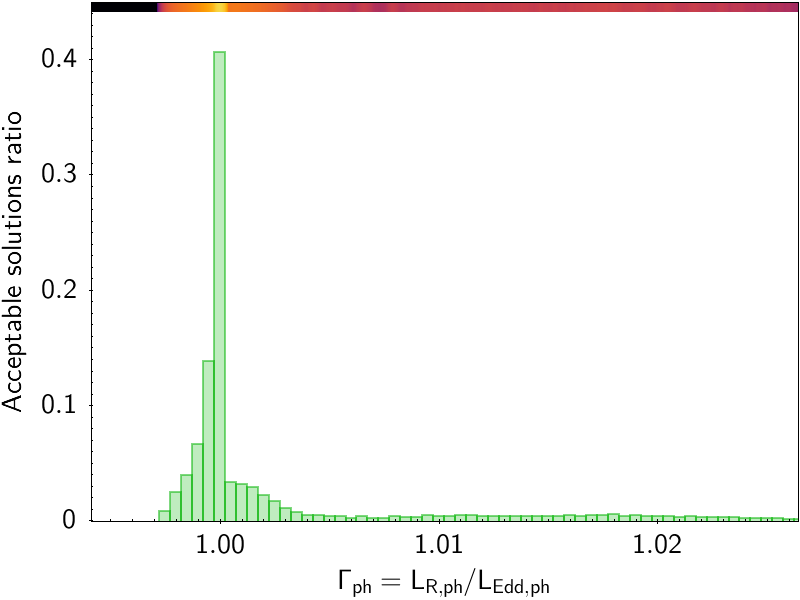}
    \caption{Distribution of photospheric luminosity ratio $\Gamma$ across acceptable solutions with varying values of model parameters.}
    \label{fig:Histogram-Photo-Gamma}
\end{figure}

A close correlation was found between several photospheric magnitudes across solutions. First, for most profiles, the photospheric luminosity was found to be almost equal to the local Eddington luminosity, as can be seen in Fig. \ref{fig:Histogram-Photo-Gamma}
 where the distribution peaks sharply around $\Gamma_\text{ph} = 1$, and the values expressed in terms of mean and standard deviation are:
$\Gamma_\text{ph}  =  1.0018 \pm 0.0055 $.
This can also be seen in the color maps of the wind parameter space in Fig. \ref{fig:WiMPS-Photo}, where the coloring for photospheric $\Gamma$ is mostly homogeneous, except for a small rise in the low-energy and high-mass outflows region where a bump in opacity table causes the local Eddington luminosity value to drop a little. The value of radiative luminosity (and therefore also opacity) at the photosphere also shows little variation across models, with the mean and standard deviation being:
$ L_\text{R,ph} / L_\text{o}   = 1.025 \pm 0.022$ and 
$ \kappa_\text{ph} / \kappa_\text{o} = 0.981 \pm 0.024$.

As both photospheric $\kappa$ and $L_\text{R}$ show very little variation, two correlations arise from the photospheric boundary conditions. Assuming a fairly constant $L_\text{R}$, the effective temperature condition gives:
\begin{align}
  &&
  r^2_\text{ph} T^4_\text{ph} \simeq \texttt{const} \Longrightarrow T_\text{ph} \sim {r_\text{ph}}^{-1/2}.
\end{align}
Similarly, for the optical depth condition and $\kappa_\text{ph} \simeq \texttt{const}$,
\begin{align}
  &&
  r_\text{ph} \, \rho_\text{ph} \simeq \texttt{const} \Longrightarrow \rho_\text{ph} \sim {r_\text{ph}}^{-1}.
\end{align}

Furthermore, by using both photospheric conditions, the equations for conservation of mass and energy, and assuming that the radiative luminosity is close to the Eddington limit, we can arrive at the following expression, which relates observable photospheric magnitudes with wind model parameters. 
\begin{align}
    &&
    \dfrac{8}{3}\ \dfrac{v_\text{ph}}{c} =  \dfrac{GM}{r_\text{ph}} \dfrac{\dot M}{L_\text{R,ph}} \simeq \dfrac{\dot E }{L_\text{R,ph}}-1  .
    \label{eq: photospheric correlations with parameters}
\end{align}
 
Such correlations can be derived considering the following. First, using mass conservation to replace $\dot M$, and assuming $L_\text{R} = L_\text{Edd} = \frac{4 \pi cGM}{\kappa}$, it is easy to see that:
\begin{align}
    &&
    \frac{\dot M}{L_\text{R}} \frac{GM}{r}= 4\pi r^2 \rho v \frac{GM}{r}\frac{ \kappa}{4\pi c GM} = \kappa \rho r \frac{v}{c}.
\end{align}
This relationship is valid whenever the luminosity is equal to the local Eddington limit. Further, particularly at the photosphere, using the optical condition, we have:
\begin{align}
    &&
    \frac{\dot M}{L_\text{R,ph}} \frac{GM}{r_\text{ph}}= \frac{8}{3} \frac{v_\text{ph}}{c}. 
    \label{Equation:Mass-cons-photo}
\end{align}
This accounts for the first equality of expression \ref{eq: photospheric correlations with parameters}.

Now let us take the energy conservation equation and rewrite it in the form:
\begin{align}
    &&
    \frac{\dot E}{L_\text{R}} - 1 = \frac{\dot M }{L_\text{R}} \frac{GM}{r} \left( \frac{v^2}{u^2} +\frac{5s^2}{u^2} + \frac{8}{3} \frac{a T^4}{\rho u^2} - 1\right),
\end{align}
where we have introduced the notation $s^2 = \frac{kT}{\mu m_A}$ for local sound speed and $u^2= \frac{2GM}{r}$ for escape velocity. For the wind to be optically thick, the critical point must lie at much greater optical depth than the photosphere. Since the critical point is the place where the wind becomes supersonic, this often translates into having $v^2 >> s^2$ at the photosphere. Lastly, in almost every solution obtained here, photospheric escape velocity is still around one order of magnitude above wind velocity, which means $s^2 << v^2 << u^2$. The last relation is relaxed as $\dot M$ increases and $\dot E$ decreases. So far, neglecting the corresponding terms we have {at the photosphere}:
\begin{align}
    &&
    \frac{\dot E}{L_\text{R}} - 1 = \frac{\dot M }{L_\text{R}} \frac{GM}{r} \left( \frac{8}{3} \frac{a T^4}{\rho u^2} - 1\right) . 
    \label{Equation:Energy-Aprox1}
\end{align}
If we now use the photospheric boundary conditions rewritten in a convenient form:
\begin{align}
    &&
    a r^2 T^4 &= \frac{L_\text{R}}{\pi c}
    & \text{and} 
    &&
    \frac{8}{3}\frac{1}{\rho r}  &= \kappa,
\end{align}
by taking the product of both and dividing by $4GM$, we get:
\begin{align}
    &&
    \frac{2}{3} \frac{aT^4 r}{\rho GM} &= \frac{\kappa L_R}{4\pi c GM} 
    & \text{or}
    && 
    \frac{4}{3} \frac{a T^4}{\rho u^2}= \frac{L_\text{R}}{L_\text{Edd}} ,
\end{align}
and if again $L_\text{R} = L_\text{Edd}$, we get at the photosphere:
\begin{align}
    &&
    \frac{8}{3} \frac{a T^4}{\rho u^2} = 2.
\end{align}
By replacing this value in equation \ref{Equation:Energy-Aprox1} we arrive finally at the relation
\begin{align}
    &&
    \frac{\dot E}{L_\text{R,ph}} - 1 \simeq \frac{\dot M }{L_\text{R,ph}} \frac{GM}{r_\text{ph}} ,
\end{align}
which is the second equality in expression \ref{eq: photospheric correlations with parameters}.

For solutions for which the approximation $v^2 << u^2$ weakens, one should add the term corresponding to wind kinetic energy $\frac{\dot M v^2}{2 L_\text{R,ph}}$:
\begin{align}
  &&
  \frac{\dot E}{L_\text{R,ph}} - 1 \simeq \frac{8}{3} \frac{v_\text{ph}}{c} +\frac{\dot M c^2}{2 L_\text{R,ph}} \left(\frac{v_\text{ph}}{c} \right)^2 , 
\end{align}
where we used relation \ref{Equation:Mass-cons-photo} to write the equation in terms of $v/c$. As $v<<c$, and photospheric luminosity varies weakly across solutions ($L_\text{R,ph} \sim L_\text{o}$), the correction term will only become relevant for relatively high values of $\dot M$. That is, if 
\begin{align}
  &&
  \frac{8}{3} \sim \frac{\dot M c v_\text{ph} }{2 L_\text{R,ph}}.
\end{align}

\section{Conclusions and Discussion} \label{sect:discuss}

In the previous section, we present several results, including a set of wind profiles (solutions for the wind model), a high-resolution parameter space exploration characterizing different physical magnitudes at points of interest, and several correlations found for photospheric (observable) values.

\begin{figure}
 \centering
 \includegraphics[keepaspectratio=true,width=8cm,clip=true,trim=0pt 0pt 0pt 0pt]{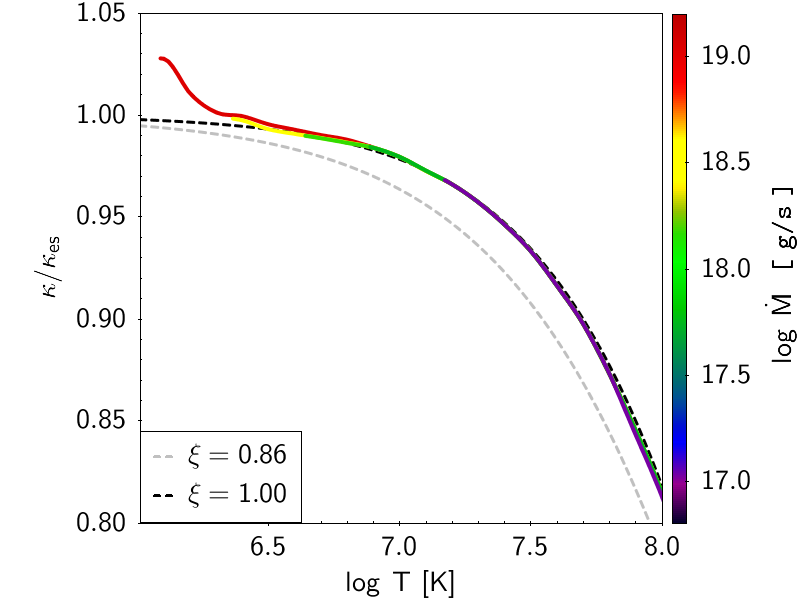}
 \caption{{Opacity profiles vs. temperature. Acceptable solutions obtained in this work with wind base at $R_{ns} = 13 \texttt{ km}$ and different values of $\log \dot M $ (continuous lines), compared to other prescriptions adopted in previous studies (dashed lines) with different values of $\xi$ (see text). Opacity is normalized to $\kappa_{es} = 0.2 \texttt{ cm}^2/\texttt{g}$.}}
 \label{fig:Opacity comparison}
\end{figure}

The wind profiles found in this work (Sect. \ref{sect: wind profiles} and Fig. \ref{fig:Profiles}) are qualitatively similar to those found in previous works (see, e.g., \cite{QuinnPacz1985}).
{These} previous studies frequently relied on an approximated expression for the opacity, in the form:
\begin{align}
    \label{Eq opacity QP} & & \kappa_{(T)}&= \kappa_\text{es} (1+X) \left[ 1 + (\alpha T )^{\xi} \right]^{-1}
\end{align}
where $\kappa_\text{es} = 0.2\ \texttt{cm}^2 \texttt{g}^{-1}$ is the electron-scattering opacity, $X$ is the hydrogen mass fraction, $ \alpha^{-1} = 4.5 \times 10^{8}\ \texttt{K} $, and $\xi$ is a parameter for which different values haven been adopted ($\xi = 1$, \cite{Kato1983,EbiHanaSugi1983}, $ \xi \simeq 0.86$, \cite{QuinnPacz1985}).
Figure \ref{fig:Opacity comparison} shows a comparison of the opacity profiles obtained for several solutions reported in this work with values obtained through the approximate Eq. \ref{Eq opacity QP} in some of the aforementioned previous studies.
Equation \ref{Eq opacity QP} neither takes into account the expected density dependence nor metallicity effects \footnote{See, however, \cite{JossMelia1984,Paczynski1983}, for other density-dependent prescriptions which still do not include metallicity effects.}. 
According to more recent studies, a high concentration of heavy elements is expected in the envelope after XRBs (see, for example, \cite{Jose2010,Woosley_2004,Fisker_2008}), and therefore high opacity. {The inclusion of updated opacity tables was a necessary improvement in this regard.}

At sufficiently high temperatures (close to the wind base), opacity drops considerably in all prescriptions, making the contribution of radiation to wind acceleration less important, while favoring the effect of gas pressure. As temperature and density drop 
outwards, the gas pressure gradient diminishes, while the opacity increases, lowering the local Eddington limit. As a result, the wind is further accelerated and sustained by radiation. This increase in opacity is higher {close to the photosphere with opacity tables} than those reported in previous studies, even for the trial $Z = 0.1$ used in this work. 
{The differences may seem relatively small here, namely on the order of a few percent, but they are expected to rise even more for higher metallicity. As we plan to study stellar winds with higher metallicities and different mixtures of heavy elements given by XRBs simulations in future work, we consider the inclusion of opacity tables to be an important step towards that goal.}

With regard to the parameter space exploration presented in this work (Sect. \ref{sect:WiMPS}, Figs. \ref{fig:WiMPS-Photo} to \ref{fig:WiMPS-Base}), it served two purposes. The first was to find the regions for which acceptable solutions exist for the wind model used, by testing part of the assumed hypothesis (i.e., optically thick and stationary wind), in a similar way to \cite{QuinnPacz1985}. The main improvement in this regard is the number of solutions obtained in the present work with respect to these latter authors. This allows for a higher resolution, which helps, for instance, to better determine these regions. The other purpose of the parameter space exploration was the characterization of physical magnitudes at points of interest, {especially} at the photosphere (high resolution enabled this too). The dependence of these photospheric values on the wind parameters serves as a link between the physics of the layers close to the base of the envelope and observable magnitudes at the photosphere in the following way: The parameters explored, namely mass and energy outflow $(\dot M, \dot E)$, are fully determined when imposing suitable boundary conditions for all physical variables at the base of the wind envelope. To this end, we would need to rely on hydrodynamic XRB models of the underlying layers, where nuclear reactions take place. We intend to do so in future work. In order for the current wind model to be applied safely and consistently, the boundary conditions provided by a given XRB model should result in mass and energy outflows that lie in the acceptable region. Once the parameters are fixed at the base by the XRB model, the corresponding wind solution gives the values of each observable magnitude at the photosphere. 
However, each value of a single photospheric magnitude taken separately does not {uniquely} determine the $(\dot M, \dot E)$ pair.

The correlations found in Sect. \ref{sect:correlations} (expression \ref{eq: photospheric correlations with parameters})
relate the photospheric magnitudes (that can either be directly observed or calculated from observable ones) with model parameters $(\dot M, \dot E)$ in a more direct and {unique} way. For instance, the radiative luminosity $L_R$ can be estimated if the distance to the XRB source is known, while wind velocity $v$ could be estimated from the Doppler shift in spectral absorption lines during the XRB.
With both velocity and luminosity, one could calculate the total energy outflow $\dot E$ by means of Eq. \ref{eq: photospheric correlations with parameters}. An estimate of photospheric temperature (e.g., by fitting a blackbody frequency distribution for the spectrum), can be used to calculate the photospheric radius by means of the $r \sim T^{-2}$ photospheric correlation or directly through the definition of effective temperature (Eq. \ref{temperature condition}). Once we have $r$, $v$, and $L_R$ at the photosphere, calculation of the mass outflow $\dot M$ is possible through Eq. \ref{Equation:Mass-cons-photo}, or from the mass conservation (Eq. \ref{Eq basic mass}) with the density given by the  $\rho \sim r^{-1}$ photospheric correlation.
Furthermore, since nuclear reactions in XRBs take place at the interface of the envelope with the neutron star core, these correlations could provide a technique for determining the radius of the neutron star indirectly by observation of the photospheric magnitudes alone.

In summary, here we explore the possible solutions of a nonrelativistic radiative wind model in a typical XRB scenario at high resolution in the parameter space $(\dot M, \dot E)$. 
The wind profiles suggest a transition from gas pressure (in the inner regions) to radiation pressure (as the wind becomes supersonic) as the main driving mechanism. 
The inclusion of updated micro-physics data with higher metallicities and higher opacities, is important here because it reinforces the role of radiation in the wind acceleration, facilitating the above-mentioned transition. 
Radiative luminosity was found to remain very close to the local Eddington limit in this regime, and at the photosphere in particular.
A high correlation was found between different observable (photospheric) magnitudes and the model parameters. 
These correlations can be used to link observational data to the physical conditions of the underlying layers where nuclear reactions take place. {However, given some of the simplifying hypothesis of the model employed (nonrelativistic regime, optically thick wind, LTE, etc.), it remains to be determined whether or not these correlations hold once more complex studies are performed. }
Hydrodynamic models of XRBs with nuclear reactions are required to impose boundary conditions at the base of the wind envelope and will determine the model parameters. 
We will leave this task for a follow-up study, where we will explore the inclusion of general relativistic effects that may become relevant in the inner layers, and additional regimes (e.g., different chemical abundance patterns, higher envelope metallicities).

\begin{acknowledgements}

The authors would like to thank Arman Aryaeipour and Manuel Linares for fruitful discussions.
This work has been partially supported by the Spanish MINECO grant AYA2017--86274--P, by the E.U. FEDER funds, and by the AGAUR/Generalitat de Catalunya grant SGR-661/2017. This article benefited from discussions within the ``ChETEC'' COST Action (CA16117).

\end{acknowledgements}

\bibliographystyle{aa.bst} % style aa.bst
\bibliography{biblio.bib}

\begin{appendix} 

\section{Critical point substitution} \label{sect:crit-substitution}

In order to avoid the numerical difficulties involved in calculating the velocity gradient at the critical point we used a substitution. The velocity gradient in the wind equations is expressed in the form of a ratio:
\begin{align}
    &&
    \frac{r}{v}\frac{dv}{dr} &= \frac{N(r,T,v)}{D(T,v)},
    \label{Eq velocity gradient as quotient}
\end{align}
with $N,D$ being the numerator and denominator which should both become zero at the critical point in order to have a regular solution.
Explicitly, the singularity condition is given by  $D(T,v) = v^2 - \frac{kT}{\mu m_A} = 0$. Hereafter, the local isothermal sound speed will be denoted by $s$, where $ s^2 = \frac{kT}{\mu m_A}$.

Consider the following function and its derivative:
\begin{align}
    &&
    y_{(x)} &= \frac{1}{2} \left( x + \frac{1}{x} \right) =  \frac{x^2 +1}{2x} \\ 
    &&
    y'_{(x)} &= \frac{1}{2}\left( 1 - \frac{1}{x^2}  \right) = \frac{x^2-1}{2x^2}.
 \end{align}
In differential form we get: 
\begin{align}
    &&
    \frac{dy}{y} &= \frac{y'}{y} dx = \frac{x^2-1}{x^2 + 1} \frac{dx}{x } .
\end{align}
 If we now set
$
 x^2 = \frac{v^2}{s^2}$, then $
 \frac{dx}{x} = \frac{dv}{v} - \frac{1}{2} \frac{dT}{T}
$
and it can be easily found that:
\begin{align}
    &&
    \frac{dy}{y} &= \frac{\frac{v^2}{s^2}-1}{\frac{v^2}{s^2}+ 1} \frac{dx}{x } = \frac{D_{(T,v)}}{v^2 +s^2} \left(  \frac{dv}{v} - \frac{1}{2} \frac{dT}{T} \right).
\end{align}
The radial gradient of this new variable $y$ then satisfies:
\begin{align}
    &&
    \frac{r}{y}\frac{d y}{d r} &= \frac{D_{(T,v)}}{v^2 +s^2} \left( \frac{r}{v}\frac{d v}{d r} - \frac{1}{2} \frac{r}{T}\frac{d T}{d r}\right),
\end{align}
and replacing expression \ref{Eq velocity gradient as quotient} for the velocity gradient, the denominator $D$ is canceled, giving:
\begin{align}
    &&
    \frac{r}{y}\frac{d y}{d r} &= \frac{1}{v^2+s^2 } \left( N_{(r,T,v)} - \frac{D_{(T,v)}}{2} \frac{r}{T}\frac{d T}{d r}\right).
\end{align}
This derivative no longer presents a singularity and can be integrated instead of the momentum equation for the new variable $y$. 

\begin{figure}
 \centering
 \resizebox{8cm}{!}{\includegraphics{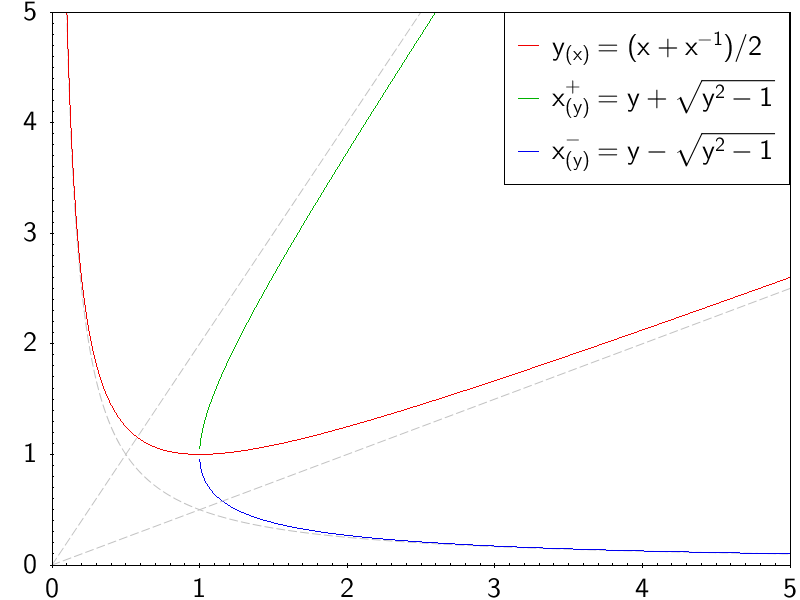}}
 \caption{Variable substitution. Change of variables $y(x)$, in red, used to integrate close to the critical point $(x,y)=(1,1)$, and the two branches of its inverse function (green and blue).}
 \label{fig:crit-substitution}
\end{figure}

However, transforming back to $v$ presents one inconvenience. The function $y(x)$ has one global minimum at $x=1$ with $y_{min}= 1$, which corresponds to the critical point (where $v=s$). From there, $y$ increases monotonically towards both $x=0$ and $x\longrightarrow \infty$ (see Fig. \ref{fig:crit-substitution}). Its inverse function therefore has two branches:
 \begin{align}
    &&
    x_{\pm} =  y \pm \sqrt{y^2 - 1},
 \end{align}
 with the following properties for all $y$:
\begin{align}
&&
 x_{+} &\geq 1 , &
 x_{-} &\leq 1 , &
 x_{-} x_{+} &= 1 , &
 x_{+} + x_{-} = 2y .
\end{align}
 Care must be taken over the choice of which branch to take.
 It is not known a priori whether at a certain point the wind should be supersonic $(x_{+})$ or subsonic $(x_{-})$, unless integration is started from the critical sonic point in a particular direction.  When integrating away from it, the value of $y$ must increase, so one must ensure that the radial gradient takes a positive value for $r \gtrsim r_{cr}$ and negative value for $r\lesssim r_{cr}$ and prevent any numerical error near the critical point that may give a different result.

\end{appendix}

\end{document}